\documentclass{article}
\usepackage{spconf,amsmath,amssymb,graphicx}
\usepackage{multirow, hhline}
\usepackage{algorithm}
\usepackage{algorithmic}

\title{Deep Long Short-Term Memory
Adaptive Beamforming Networks for Multichannel Robust Speech Recognition}
\name{Zhong Meng$^{1,2}$\sthanks{Zhong Meng performed the work while he was an intern at Mitsubishi Electric Research Laboratories,
Cambridge, MA.}, Shinji Watanabe$^{1}$, John R. Hershey$^{1}$, Hakan Erdogan$^{3}$
}
\address{$^{1}$ Mitsubishi Electric Research Laboratories, Cambridge, MA
\\ $^{2}$ Georgia Institute of Technology, Atlanta, GA
\\ $^{3}$ Microsoft Research, Redmond, WA
} 

\begin{document}
\ninept

\maketitle
\begin{abstract}
Far-field speech recognition in noisy and reverberant conditions remains a challenging
problem despite recent deep learning breakthroughs.  
This problem is commonly addressed by acquiring a
speech signal from multiple microphones and performing beamforming over them.
In this paper, we propose to use a recurrent neural network with long
short-term memory (LSTM) architecture to adaptively estimate real-time
beamforming filter coefficients to cope with non-stationary environmental
noise and dynamic nature of source and microphones positions which
results in a set of time-varying room impulse responses. The LSTM
adaptive beamformer is jointly trained with a deep LSTM acoustic model to
predict senone labels. Further, we use hidden units in the deep LSTM
acoustic model to assist in predicting the beamforming filter
coefficients. The proposed system achieves 7.97\% absolute gain over
baseline systems with no beamforming on CHiME-3 real evaluation set. 
\end{abstract}
\section{Introduction}
\label{sec:intro}
Although extraordinary performance has been achieved in automatic speech
recognition (ASR) with the advent of deep neural networks (DNNs)
\cite{dnn_hinton, dnn_dong}, the performance still degrades dramatically in
noisy and far-field situations \cite{reverb_delcroix, chime3_hori}.
To achieve robust speech recognition, multiple microphones can be used
to enhance the speech signal, reduce the effects of noise and
reverberation, and improve the ASR performance. 
In this scenario,  an essential step of the ASR front-end processing
is multichannel filtering, or \emph{beamforming},  
which steers a spatial sensitivity region, or ``beam,'' in the direction of the target source, and
inserts spatial suppression regions, or ``nulls,'' in the directions corresponding to noise and other
interference.

Delay-and-sum (DAS) beamforming is widely used for multichannel signal
processing \cite{das}, in which the multichannel inputs of an microphone
array are delayed to be aligned in time and then summed up to be a single
channel signal.  The signal from the target direction is enhanced and the
noises and interferences coming from other directions are attenuated.
Filter-and-sum beamforming applies filters to the input
channels before summing them up \cite{fas}. Minimum variance
distortionless response (MVDR) \cite{blstm_mvdr} and generalized
eigenvalue (GEV) \cite{gev_ernst} are filter-and-sum beamforming methods
which solve for filter coefficients using different derivations.

Although these methods have achieved good performance in beamforming,
their goal is to optimize only the signal-level objective (e.g., SNR). In
order to achieve robust speech recognition, it is more important to
jointly optimize beamforming and acoustic model with the
objective of maximizing the ASR performance. In \cite{bf_dnn_shinji}, the
parameters of a frequency-domain beamformer are first estimated by a DNN
based on the generalized cross correlation between microphones.
Conventional features are extracted from the beamformed signal before
passing through a second DNN for acoustic modeling. Instead of
filtering in the frequency domain, \cite{cldnn_tara} performs spatial and
spectral filtering through time-domain convolution over raw waveform. The
output feature is then passed to a convolutional LSTM DNN (CLDNN)
acoustic model to predict the context-dependent state output targets. In
\cite{cldnn_factor_tara}, the beamforming and frequency decomposition are
factorized into separate layers in the network. These approaches assume
that the speaker position and the environment are fixed and estimate
constant filter coefficients for either beamforming or spatial and
spectral filtering. 

However, in real noisy and far-field scenarios, as the position of the
source (speaker), noise and room impulse response keep changing, the
time-invariant filter coefficients estimated by these neural networks may fail to 
robustly enhance the target signal. Therefore, we
propose to adaptively estimate the beamforming filter coefficients at
each time frame using an LSTM to deal with any possible changes of the
source, noise or channel conditions. The enhanced signal is generated by
applying these time-variant filter coefficients to the short-time Fourier
transform (STFT) of the array signals. Log filter-bank like features are
obtained from the enhanced signal and then passed to a deep LSTM acoustic
model to predict the senone posterior. The LSTM beamforming network and
the LSTM acoustic model are jointly trained using truncated
back-propagation through time (BPTT) with a cross-entropy objective. 
STFT coefficients of the array signals are used as the input of the
beamforming network. In ASR systems of \cite{nmf_enhan_asr,
lstm_enhan_asr}, the speech signal is enhanced by NMF and LSTM before fed
into the acoustic model.  But speech enhancement module and the acoustic
model are not jointly optimized to minimize the WER and the input is only
single channel signal.

Previous work \cite{lstm_separation_hakan} has shown that the speech
separation performance can be improved by incorporating the speech
recognition alignment information within the speech enhancement
framework.  Inspired by this, we feed the units of the top hidden layer
of the LSTM acoustic model at the previous time step back as an auxiliary
input to the beamforming network to predict the current filter
coefficients. Note that our work is different from \cite{lstm_beamforming_bo} in that: (1) we perform adaptive beamforming over 5 different input channels, but their system works only on 2 input channels; (2) our adaptive LSTM beamformer predicts only the frequency domain filter coefficients and performs frequency domain filter-and-sum over STFT coefficients, while their work majorly focuses on the time-domain filtering with raw waveforms as the input; 
(3) the log Mel filter bank like features are generated with fixed log Mel transform over the beamformed STFT coefficients for acoustic modeling in our work, while time/frequency domain convolution is performed with trainable parameters on the beamformed features in their work; 
(4) no additional gate modulation is applied to the feedback to reduce the system complexity for our much smaller dataset.
In the experiments, we show that this feedback captures high-level knowledge about the acoustic states and increases the performance. 
The experiments are conducted with the CHiME 3 dataset. The joint training of LSTM adaptive beamforming network and deep LSTM acoustic model achieves 7.75\% absolute gain over the single channel signal on the real test data. 
The acoustic model feedback
provides an extra gain of 0.22\%.

\section{LSTM Adaptive Beamforming}
\label{sec:lstm_beamformer}

\subsection{Adaptive Filter-and-Sum Beamforming}
As a generalization of the delay-and-sum beamforming, filter-and-sum
beamformer processes the signal from each microphone using a finite impulse response (FIR) filter before summing them up. 
In frequency domain, this operation can be written as:
\begin{align}
	\hat{x}_{t,f}=\sum_{m=1}^M g_{f,m} x_{t,f,m},
	\label{eqn:filter_and_sum}
\end{align}
where $x_{t,f,m}\in \mathcal{C}$ is the complex STFT coefficient for the
time-frequency index $(t,f)$ of the signal from channel $m $, $g_{f,m}\in
\mathcal{C}$ is the beamforming filter coefficient and $\hat{x}_{t,f} \in
\mathcal{C}$ is the complex STFT coefficient of the enhanced signal.
In Eq. \eqref{eqn:filter_and_sum}, $t=1,\ldots,T, f=1,\ldots,F$ and $M, T, F$
are the numbers of microphones, time frames and frequencies.
To cope with the time-variant source position and room impulse response,
we make the filter coefficients time-dependent and propose the
adaptive filter-and-sum beamforming:
\begin{align}
	\hat{x}_{t,f}=\sum_{m=1}^M g_{t,f,m} x_{t, f, m},
	\label{eqn:filter_and_sum_adaptive}
\end{align}
where $g_{t, f, m}\in \mathcal{C}$ is time-variant complex filter
coefficient.

\subsection{Adaptive LSTM Beamforming Network}
\label{sec:lstm_bf}
The LSTM network is a special kind of recurrent neural network (RNN) with
purpose-built memory cells to store information \cite{lstm_schmid}. The
LSTM has been successfully applied to many different tasks
\cite{lstm_speech_graves, lstm_lm} due to
its strong capability of learning long-term dependencies.
The LSTM takes in an input sequence $x=\{x_1,\ldots, x_T\}$ and computes the
hidden vector sequence $h=\{h_1, \ldots, h_T\}$ by iterating the equation
below
\begin{align}
	h_t=\text{LSTM}(x_t, h_{t-1})
	\label{eqn:rnn_iter}
\end{align}
We implement the LSTM in Eq. \eqref{eqn:rnn_iter} with no
peep hole connections.

In this work, we apply {\it real-value} LSTM to the adaptive filter-and-sum beamformer to predict the real and imaginary parts of the complex filter coefficients at time $t$ and channel $m$.
That is, we introduce the following real-value vectors for complex values $g_{t,f,m}$ and $x_{t,f,m}$ in Eq.~\eqref{eqn:filter_and_sum_adaptive}:
\begin{align*}
 g_{t,m} & \triangleq \left[\Re(g_{t,f,m}), \Im(g_{t,f,m}) \right] _{f=1} ^F \in \mathcal{R}^{2F} \\
 x_{t} & \triangleq \left[ \Re(x_{t,f,m}), \Im(x_{t,f,m}) \right] _{f=1, m=1} ^{F, M} \in \mathcal{R}^{2FM}.
\end{align*}
With this representation, the real-value LSTM predicts $g_{t,m}$ as follows:
\begin{align}
	p_t&=W_{x,p}x_t \label{eqn:lstm_p} \\
	h_t&=\text{LSTM}^{BF}(p_t, h_{t-1}) \label{eqn:lstm_bf} \\
	g_{t,m}&=\tanh(W_{h,m}h_t), \quad m=1,\ldots,M  \label{eqn:lstm_filter},
\end{align}
where $W_{x,p}$ and $W_{h,m}$ are projection matrices.
We use $\tanh(\cdot)$ function to limit the range of the filter coefficients
within $[-1,1]$.


The real and imaginary parts of the STFT coefficient $\hat{x}_{t,f}$ of the beamformed signal are generated by
Eq.~\eqref{eqn:filter_and_sum_adaptive} as follows
\begin{equation}
\begin{cases} 
    \Re(\hat{x}_{t,f}) \hspace{-2.5mm}  &=\sum_{m=1}^M \Re(x_{t,f,m}) \Re(g_{t,f,m}) -
	\Im(x_{t,f,m}) \Im(g_{t,f,m}) \\ 
	\Im(\hat{x}_{t,f})&=\sum_{m=1}^M
	\Re(x_{t,f,m}) \Im(g_{t,f,m}) + \Im(x_{t,f,m}) \Re(g_{t,f,m}).
	\label{eqn:implement_beamform}
\end{cases}
\end{equation}
More sophisticated features can be extracted from the beamformed STFT
coefficients and are passed to the LSTM acoustic model to predict the
senone posterior. In our experiments, the log Mel filterbank like feature
is generated from Eq.~\eqref{eqn:implement_beamform} by
\begin{align}
	z_t&=\log\left( \text{Mel} ( P_t ) \right) \label{eqn:mel} \\
	P_t&=\left [ \Re(\hat{x}_{t,f})^2+ \Im(\hat{x}_{t,f})^2 \right ] _{f=1} ^F \in \mathcal{R} ^{F}
	\label{eqn:mag}
\end{align}
where $\text{Mel}(\cdot)$ is the operation of Mel matrix multiplication, and $P_t$ is $F$ dimensional real-value vector of the power spectrum of the beamformed signal at time $t$. 
Global mean and variance normalization is applied to this log Mel filterbank like feature. 
Note that all operations in this section are performed with the {\it
real-value} computation, and can be easily represented by a differentiable computational graph.


\subsection{Deep LSTM Acoustic Model}
\label{sec:lstm_am}
Recently, LSTMs are shown to be more effective than DNNs
\cite{dnn_hinton, dnn_dong} and conventional RNNs \cite{rnn_chao,
rnn_povey} for acoustic modeling as they are
able to model temporal sequences and long-range dependencies more
accurately than the others especially when the amount of training data is
large. LSTM has been successfully applied in both the RNN-HMM hybrid
systems \cite{lstm_hmm_graves, lstm_hmm_sak} and the end-to-end system
\cite{lstm_ctc_graves1, lstm_attention_chorowski1}.

In this work, the deep LSTM-HMM hybrid system is utilized for acoustic
modeling. A forced alignment is first generated by a GMM-HMM system and
is then used as the frame-level acoustic targets which the LSTM attempts to
classify. The LSTM is trained with cross-entropy objective function using
truncated BPTT. 
In this paper, to connect the deep LSTM with the adaptive LSTM
beamformer, we compute log Mel filterbank $z_t$ from the beamformed STFT
coefficients.
\begin{align}
	q_t&=W_{z,p}z_t \label{eqn:lstm_am_p} \\
	s_t&=\text{LSTM}^{AM}(q_t, s_{t-1}) \label{eqn:lstm_am} \\
	y_t&=\text{softmax}(W_{s,y}s_t) \label{eqn:lstm_softmax}
\end{align}
$q_t$ is the projection of $z_t$ into a high-dimensional space and $y_t$
is the senone posterior.  

\subsection{Integrated Network of LSTM Adaptive Beamformer and Deep LSTM Acoustic Model}
\label{sec:joint}
\begin{figure}[htpb!]
	\centering
	\includegraphics[width=0.8\columnwidth]{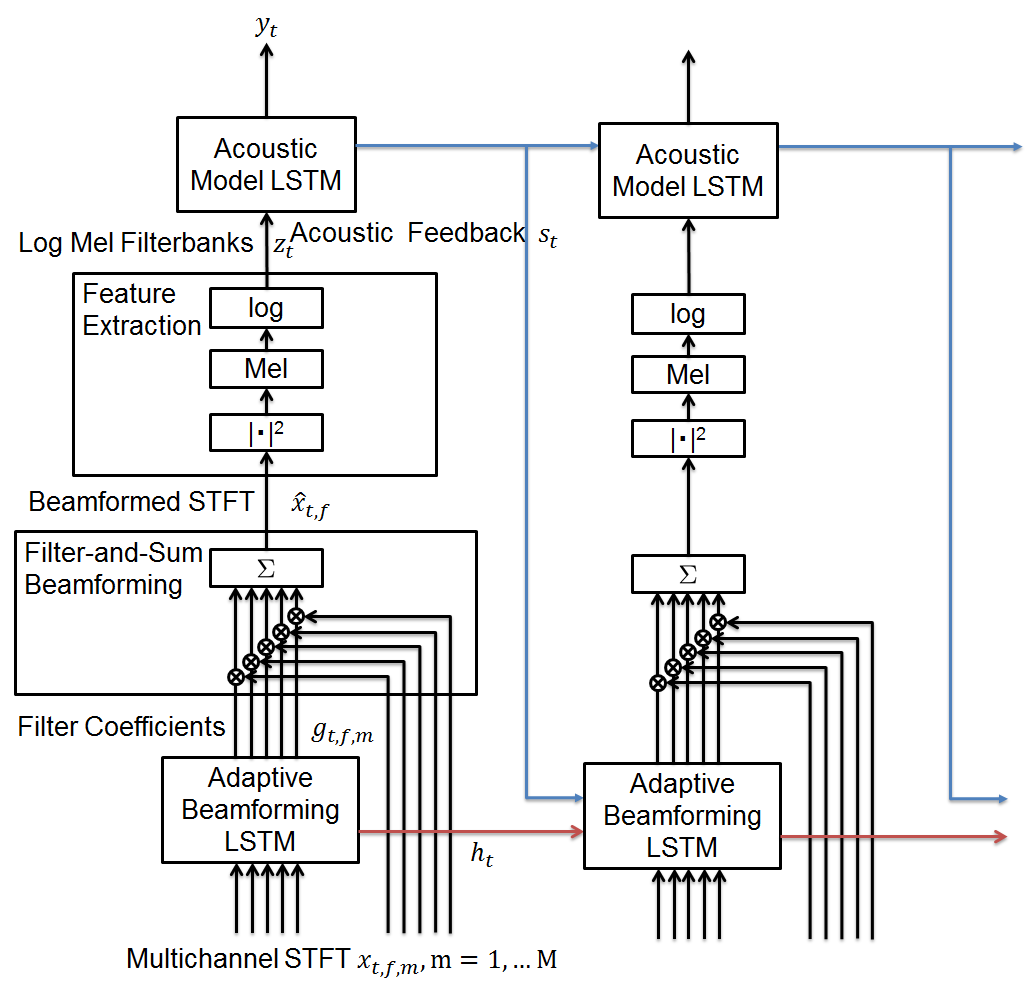}
	\caption{The unfolded integrated network of an LSTM adaptive beamformer and an LSTM
	acoustic model. The acoustic feedback (in blue) is introduced to
allow the hidden units in LSTM acoustic model to assist in predicting the
filter coefficient at current time.} \label{fig:joint_lstm_asrfeedback}
\end{figure}

In order to achieve robust speech recognition by making use of
multichannel speech signals, LSTM beamformer in Section \ref{sec:lstm_bf}
and the deep LSTM acoustic model in Section \ref{sec:lstm_am} need to be
jointly optimized with the objective of maximizing the ASR performance.
In other words, the beamforming LSTM needs to be concatenated with the
LSTM acoustic model to form an integrated network that takes multichannel
STFT coefficients as the input and produces senone posteriors as
illustrated in Fig. \ref{fig:joint_lstm_asrfeedback}. The deep LSTM has
three hidden layers in our experiments but only one is shown here for
simplicity. 

To train the integrated LSTM network, we connect the beamforming network
\eqref{eqn:filter_and_sum_adaptive} -- \eqref{eqn:lstm_filter}, log Mel
filtering \eqref{eqn:mel}, and the acoustic model
\eqref{eqn:lstm_am_p} -- \eqref{eqn:lstm_softmax} as a single feed forward
network, and back-propagate the gradient of the cross-entropy objective
function through the network so that both the adaptive beamformer and the
acoustic model are optimized for the ASR task by using multi-channel
training data.

On top of that, we feed the hidden units
of the top hidden layer of the deep LSTM acoustic model back to the input of
the LSTM beamformer as the auxiliary feature to predict the filter
coefficients at next time. By introducing the acoustic model
feedback, the Eq. \eqref{eqn:lstm_bf} is re-written as
\begin{align}
	h_t&=\text{LSTM}^{BF}((p_t, s_{t-1}), h_{t-1})
\end{align}
where $(p_t, s_{t-1})$ is the concatenation of the acoustic feedback from
previous time $s_{t-1}$ and the current projection $p_t$.

Direct training of the integrated network easily falls into a local
optimum as the gradients for the LSTM beamformer and the deep LSTM acoustic
model have different dynamic ranges. For a robust estimation of the model
parameters, the training should be performed in sequence as shown in Algorithm \ref{alg:train_steps}. 

\begin{algorithm}
\caption{Train LSTM adaptive beamformer and deep LSTM acoustic model}
\label{alg:train_steps}
\begin{algorithmic}[1]
\STATE \label{step:am} Train a deep LSTM acoustic model with log Mel
filterbank feature extracted from the speech of all channels to minimize
the cross-entropy objective.
\STATE Initialize the integrated network with the deep LSTM acoustic model in Step
	\ref{step:am}.
\STATE \label{step:semi_joint} Train the integrated network with the ASR
cross-entropy objective, update only the parameters in the LSTM
beamformer.
\STATE \label{step:full_joint} Jointly train the integrated network in
	Step \ref{step:semi_joint} with the ASR cross-entropy objective,
updating all parameters in the LSTM beamformer and deep LSTM acoustic model.
\STATE \label{step:asrfeedback_joint} Introduce the acoustic feedback and
	re-train the integrated network with the ASR objective, updating
	all the parameters. 
\end{algorithmic}
\end{algorithm}


\section{Experiments}
\label{sec:exp}
\subsection{CHiME-3 Dataset}
The CHiME-3 dataset is released with the 3rd CHiME speech Separation and
Recognition Challenge \cite{chime3_barker}, which incorporates the Wall
Street Journal corpus sentences spoken by talkers situated in challenging
noisy environment recorded using a 6-channel tablet based microphone
array.
CHiME-3 dataset consists of both real and simulated data. The real data
is recorded speech spoken by actual talkers in four real noisy
environments (on buses, in caf\'{e}s, in pedestrian areas, and at street
junctions). To generate the simulated data, the clean speech is first
convoluted with the estimated impulse response of the environment and
then mixed with the background noise separately recorded in that
environment \cite{chime3_hori}. The training set consists of 1600 real
noisy utterances from 4 speakers, and 7138 simulated noisy utterances
from the 83 speakers in the WSJ0 SI-84 training set recorded in 4 noisy
environments. There are 3280 utterances in the development set including
410 real and 410 simulated utterances for each of the 4 environments.
There are 2640 utterances in the test set including 330 real and 330
simulated utterances for each of the 4 environments. The speakers in
training set, development set and the test set are mutually different
(i.e., 12 different speakers in the CHiME-3 dataset). The training,
development and test data are all recorded in 6 different channels. The WSJ0
text corpus is also used to train the language model. 

\subsection{Baseline System}
\label{sec:exp_baseline}
The baseline system is built with Chainer \cite{chainer} and Kaldi
\cite{kaldi} toolkits.  40-dimensional log Mel filterbank features
extracted by Kaldi from all 6 channels are used to train a deep LSTM acoustic
model using Chainer. The LSTM has 3 layers and each hidden layer has 1024
units. The output layer has 1985 units, each of which corresponds to a
senone target. The input feature is first projected to a 1024 dimensional
space before being fed into the LSTM. The forced alignment generated by a
GMM-HMM system trained with data from all 6 channels is used as the
target for LSTM training.  During evaluation, only the development and
test data from the $5^{\text{th}}$ channel is used for testing (only for
the baseline system). The LSTM is
trained using BPTT with a truncation size of $100$ and a learning rate of $0.01$.
The batch size for stochastic gradient descent (SGD) is 100. The WER
performance of the baseline system is shown in Table \ref{table:wer}.

\begin{table}
\centering
\begin{tabular}[c]{c|c|c|c|c|c}
	\hline
	\hline
	\multirow{2}{*}{\footnotesize\begin{tabular}{@{}c@{}}
		System \end{tabular}} &
	\multirow{2}{*}{\footnotesize\begin{tabular}{@{}c@{}}Input \\
		Feature \end{tabular}} &
	\multirow{2}{*}{\footnotesize\begin{tabular}{@{}c@{}} Simu \\ Dev
	\end{tabular}} & 
	\multirow{2}{*}{\footnotesize\begin{tabular}{@{}c@{}} Real \\ Dev
	\end{tabular}} & 
	\multirow{2}{*}{\footnotesize\begin{tabular}{@{}c@{}} Simu \\ Test 
	\end{tabular}} & 
	\multirow{2}{*}{\footnotesize\begin{tabular}{@{}c@{}} Real \\ Test 
	\end{tabular}} \\
	&&&&& \\
	\hline
        AM (baseline) & Fbank & 16.15 & 19.24 & 23.02 & 32.88 \\
	\hline
	BeamformIt+AM & STFT & 14.32 & 12.99 & 24.36 & 21.21 \\
	\hline
	BF+AM (fixed) & STFT & 15.23 & 15.01 & 23.14 & 25.64
	\\
	\hline
	BF+AM & STFT & 14.43 & 15.19 & 22.40 & 25.13
	\\
	\hline
	BF+AM+Feedback & STFT & 14.28 & 15.10 & 22.23 & 24.91 \\
	\hline
	\hline
\end{tabular}
  \caption{The WER performance (\%) of the baseline LSTM acoustic model
	  (AM), BeamformIt-enhanced signal as the input of the AM, 
	  joint training of LSTM beamformer and LSTM acoustic model
	  (BF+AM) with or without acoustic feedback.} 
	\label{table:wer}
\end{table}

\subsection{LSTM Adaptive Beamformer}
\label{sec:exp_bf}
The 257-dimensional complex STFT coefficients are extracted for the
speech in channels $1, 3, 4, 5, 6$. The real and imaginary parts of STFT
coefficients from all the 5 channels are concatenated together to form $257 \times 2 \times 5=2570$ dimensional input of the beamforming
LSTM.  The input is projected to 1024 dimensional space before being fed
into the LSTM. The beamforming LSTM has one hidden layer with 1024
units. The hidden units vector is projected to 5 sets of
$257\times2=514$ dimensional filter coefficients for adaptively
beamforming signals from 5 channels using Eq.
\eqref{eqn:filter_and_sum_adaptive}. The MSE objective is computed
between the beamformed signal and BeamformIt \cite{beamformit}. The
beamforming LSTM is trained using BPTT with a truncation size of $100$, 
a batch size of $100$ and a learning rate of $1.0$.

\subsection{Joint Training of the Integrated Network}
\label{sec:exp_joint}
The baseline LSTM acoustic model trained in Section
\ref{sec:exp_baseline} and the LSTM adaptive beamformer trained in
Section \ref{sec:exp_bf} are concatenated together as the
initialization of the integrated network. A feature extraction layer is
inserted in between the two LSTMs to extract 40-dimensional log Mel
filterbank features with Eq. \eqref{eqn:mel}.  The integrated network is
trained in a way described in Steps \ref{step:semi_joint},
\ref{step:full_joint} and \ref{step:asrfeedback_joint} of Section
\ref{sec:joint}.  BPTT with a truncation size of $100$ and a batch size
of $100$ and a learning rate of $0.01$ is used for training. The data from
all $5$ channels in the development and test set is used for evaluating
the integrated network. The WER performance for different cases are shown
in Table \ref{table:wer}. 

\subsection{Result Analysis}
\label{sec:exp_result}
From Table \ref{table:wer}, the best system is the integrated network of
an LSTM adaptive beamformer and a deep LSTM acoustic model with the
acoustic feedback, which achieves 14.28\%, 15.10\%, 22.23\%, 24.91\% WERs
on the simulated development set, real development set, simulated test
set and real test set of the CHiME-3 dataset respectively. The joint
training of the integrated network without updating the deep LSTM
acoustic model achieves absolute gains of 0.92\%, 4.23\% and 7.24\% over
the baseline system on the simulated development set, real development
set and real test set respectively. The joint training of the integrated
network with all the parameters updated achieves absolute gains of
1.72\%, 4.05\%, 0.62\% and 7.75\% respectively over the baseline systems
on the simulated development set, real development set, simulated test
set and real test set respectively. The large performance improvement
justifies that the LSTM adaptive beamformer is able to estimate the
real-time filter coefficients adaptively in response to the changing
source position, environmental noise and room impulse response with the
LSTM acoustic model jointly trained to optimize the ASR objective.
Further absolute gains of 0.15\%, 0.09\%, 0.17\% and 0.22\% are achieved
with the introduction of acoustic feedback, which indicates that the
high-level acoustic information is also helpful in predicting the filter
coefficients at the next time step. 

Note that although the proposed system with acoustic feedback achieves
0.04\% and 2.13\% absolute gains over the beamformed signal generated by
BeamformIt on the simulated development and test sets, it does not work
as well as the BeamformIt on the real development and test sets. In
BeamformIt implementation, the two-step time delay of arrival Viterbi
postprocessing makes use of both the past and future information in
predicting the best alignment of multiple channels at the current time,
while in our system, only the history in the past is utilized to estimate
the current filter coefficients. This may explain the differences in WER
performance and can be alleviated by using bidirectional LSTM as
part of the future work.

\subsection{Beamformed Feature}
\begin{figure}[htpb!]
	\centering
	\includegraphics[width=1.0\columnwidth]{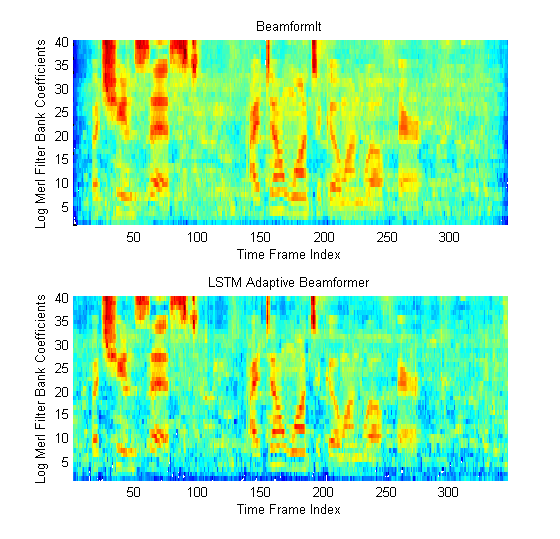}
	\caption{The comparison of the log Mel filter bank coefficients of
		the same utterance extracted from
	STFT coefficients beamformed by BeamformIt (upper) and LSTM adaptive beamformer
(lower) .}
	\label{fig:logmel_fbank}
\end{figure}

The LSTM beamformer adaptively predicts the time-variant beamforming
coefficients and performs filter-and-sum beamforming over the 5 input
channels. The log Mel filter bank feature is obtained from the STFT
coefficients. From Fig. \ref{fig:logmel_fbank}, we see that the log Mel
filter bank feature obtained from the LSTM adaptive beamformer is
quite similar to the log Mel filter bank feature extracted from the STFT
coefficients beamformed by BeamformIt for the same utterance. The SNR is
not high but matches the LSTM acoustic model well for maximizing the ASR
performance.

\section{Conclusions}
\label{sec:conclusions}

In this work, LSTM adaptive beamforming is proposed to adaptively predict
the real-time beamforming filter coefficients to deal with the time-variant
source location, environmental noise and room impulse response inherent in
the multichannel speech signal. To achieve robust ASR, the LSTM adaptive
beamformer is jointly trained with a deep LSTM acoustic model to optimize
the ASR objective. This framework achieves absolute gains of 1.72\%,
4.05\%, 0.62\% and 7.75\% over the baseline system on the CHiME-3 dataset.
Further improvement is achieved by introducing the acoustic feedback to
assist in predicting the filter coefficients. However, our approach does
not work as well as the BeamformIt on real data and we will look into this
in the future.

\vfill\pagebreak


\bibliographystyle{IEEEbib}
\bibliography{refs}

\end{document}